\documentclass{iaus}
\usepackage{graphicx}
\usepackage{index}
\newindex{author}{idx}{ind}{Author Index}
\newindex{object}{odx}{ond}{Object Index}
\newindex{subject}{sdx}{snd}{Subject Index}
\makeindex

\newcommand{\ltsima}{$\; \buildrel < \over \sim \;$}
\newcommand{\simlt}{\lower.5ex\hbox{\ltsima}}
\newcommand{\gtsima}{$\; \buildrel > \over \sim \;$}
\newcommand{\simgt}{\lower.5ex\hbox{\gtsima}}

\title[The SFH of the Milky Way Galaxy] 
{The star-formation history of the Milky Way Galaxy}

\author[Rosemary F.G.~Wyse]   
{Rosemary F.G.~Wyse}
\index[author]{Wyse, R.F.G.}

\affiliation{Department of Physics \& Astronomy, Johns Hopkins University, \\ Baltimore, MD 21218, USA  \\ email: {\tt wyse@pha.jhu.edu}}

\pubyear{2009}
\volume{258}  
\pagerange{119--126}
\jname{The Ages of Stars }
\editors{E.E. Mamajuk, D.R. Soderblom,  \& R.F.G. Wyse, eds.}
\begin{document}

\maketitle

\begin{abstract}
The star-formation histories of the main stellar components of the
Milky Way constrain critical aspects of galaxy formation and
evolution. I discuss recent determinations of such histories, together
with their interpretation in terms of theories of disk galaxy
evolution.


\index[subject]{Galaxy: disk}
\index[subject]{Galaxy: evolution}
\index[subject]{Galaxy: stellar content}
\end{abstract}

\firstsection 
\section{Context: Disk galaxy formation and evolution}

The Milky Way appears to be a typical disk Galaxy, albeit the one for
which we can obtain the most detailed information, for the largest
samples of tracer objects (I will mostly discuss samples of stars).
The star-formation histories of the main stellar components of the
Milky Way constrain critical aspects of disk galaxy formation and
evolution. These include aspects of the merger history, such as what
merged with the Milky Way and when did it merge, and also the epoch at
which extended disks started to form.  These in turn depend upon the
nature of dark matter, possible `feedback' mechanisms once stars start
to form, and the amplitudes, onset and duration, and signs, of gas
flows.

Much observational evidence, in particular from large-scale structure
such as the spectrum of fluctuations in the cosmic microwave
background, and the statistics of massive galaxy clusters, has led to
a `concordance' cosmological model, wherein the recent (since a
redshift of $\sim 1$) rate of overall expansion of the Universe is
driven by Dark Energy, and the matter content of the Universe is
predominantly non-baryonic Cold Dark Matter (e.g.~Spergel et
al.~2007).  The primordial power spectrum of Cold Dark Matter (CDM) has most
power on small scales, resulting in a hierarchical sequence of
structure formation, whereby small scales form first, then
subsequently merge to form larger systems.  This concordance model
will be referred to below as $\Lambda$CDM.

While baryons are a minor constituent of such a $\Lambda$CDM Universe,
contributing less than 3\% of the energy density at the present day,
they are of course how we trace most of the mass in the Universe.  The
baryonic physics of gaseous dissipational cooling and subsequent star
formation is much more difficult to model than is the (Newtonian)
gravity that is the only force of significance for dissipationless
CDM.  The most detailed simulations of the formation and
evolution of an analog of the present-day Milky Way have therefore
been purely N-body, with gravity the only force and all matter being CDM.  The state-of-the art
in October~2008 is illustrated by the Via~Lactae~II simulation
(Diemand et al.~2008). This simulation follows the formation of the
dark halo of a model Milky Way Galaxy, with initial conditions
selected from a larger simulation to ensure that there will be no
major merger after a redshift of 1.7 (a look-back time of somewhat
less than 10~Gyr -- as we discuss below, this restriction on the merger
history is to prevent destruction of the thin disk within the lifetime
of old stars presently observed in the local disk).

A striking feature of this very high-resolution simulation is the
persistent substructure, even near the analog of the solar circle: 97\%
of subhaloes that are initially identified at redshift unity still
leave a bound remnant at the present epoch.  Depending on the details
of their mass and density distributions, these subhaloes are expected
to lead to heating and fattening of the thin stellar disk
(e.g.~Hayashi \& Chiba 2006).
 
The overall trends of stellar ages in such hierarchical clustering
models are illustrated by the predictions of Abadi et al.~(2003), who
simulated the formation of a disk galaxy using a hybrid N-body and
Smoothed Particle Hydrodynamics code, with simple gas-cooling and
star-formation criteria.\footnote{Watch the movie at
\tt{http://www.aip.de/People/MSteinmetz/Movies.html}} The galaxy that
forms is more bulge-dominated than is the Milky Way, reflecting the
typical, active merger history in the $\Lambda$CDM cosmogony (and
hence the need to pre-select initial conditions to form a Milky Way
analog, as done by Diemand et al.).  The model galaxy can be
decomposed into spheroid (bulge/halo), thick disk and thin disk, with
these components being distinct in terms of both surface-brightness
profiles and kinematics. The spheroid is old, and was created by major
mergers that disperse the pre-existing disk, plus tidal debris from
more minor mergers (in this simulation, about half by each
mechanism). The thin disk seen today is mostly stars that formed {\it
in situ\/} after the cessation of merger activity, from gas that was
accreted smoothly into the disk plane, as it cooled in the dark halo.
The inner thin-disk formed faster, reflecting the shorter accretion
times of lower-angular-momentum gas, destined to dissipate into
circular orbits in the central regions (Mo, Mao \& White 1998). There
is therefore a gradient in mean stellar age at the present epoch,
declining from a mass-weighted age of 6~Gyr close to the center to
3~Gyr in the outer parts.  The active merging, typical for the $\Lambda$CDM cosmogony, 
means that the overall galactic potential is not steady enough for
formation of an extended thin disk  until more recently than a look-back time
of $\sim 8$~Gyr (redshifts less than unity). A disk can form early, then be destroyed, but the early disks are compact, due to  the low angular-momentum content of their progenitor gas, reflected in the `inside-out' formation of disks  (e.g.~Scannapieco et al.~2009).

In the Abadi et al.~realization,  the thick disk
is predominantly old, and the bulk of it consists not of stars formed
in the potential well of the final galaxy, but instead of stars formed
in satellite galaxies that were later accreted, with half due to a
satellite that merged $\sim 6$~Gyr ago, while the stars are older than
$\sim 10$~Gyr old.  The old stars (older than $\sim 8$~Gyr) in the
thin disk are also predominantly accreted.  This of course requires
that the satellite hosts be on high-angular-momentum  orbits, with periGalacticons within the disk, before their
stars are assimilated into the disk.  Given that the typical initial orbit of a satellite dark-halo is far from circular, this in turn requires that the
satellite halo be rather massive, so that dynamical friction can
operate to circularize and shrink the orbit.
 
\subsection{Tracing the  merger history} 
 
In a merger, orbital energy is absorbed into the internal
degrees of freedom of the merging systems, heating them.  A major
merger, one with approximately equal mass ratio, essentially destroys
pre-existing disks, transforming them into a pressure-supported
spheroid or stellar halo (with $\sim 1:3$ being the limiting mass
ratio; Cretton et al.~2001).  The effects of the accretion of
satellites (galaxies plus pure dark matter) depends on the time of
accretion, their initial orbits, masses and density profiles, since
these dictate how easily they are tidally disrupted, and to which
component their stellar debris will contribute.  Minor mergers (mass
ratios of less than $\sim 1:5$) of a fairly robust satellite should
puff up an existing thin stellar disk into a thick disk (e.g.~Quinn \&
Goodman 1986; Kazantzidis et al.~2008), plus deposit tidal debris from
the satellite along its orbit.  Some fraction of the stellar mass of
the thin disk is also often assumed, in semi-analytic models, to be
directly added to the bulge after a minor merger (Kauffmann 1996; de
Lucia \& Blaizot 2007). Gravitational torques during mergers cause
transport of angular momentum, resulting in gas being taken into the
central regions, where, after an induced star-burst, it can contribute
to the bulge (Mihos \& Hernquist 1996). Gas flows driven by mergers
can possibly also build-up the bulge by triggering disk instabilities,
in a fusion of dynamical and secular mechanisms (Bower et al.~2006).
Dense, inner regions of massive satellites (the higher mass meaning
dynamical friction timescales are shorter) could also contribute stars
to the bulge (Ostriker \& Tremaine 1975).

The stellar age distributions of thin disk, thick disk, stellar halo
and bulge populations therefore depend on the merger history, and 
observational determination of these distributions can constrain the
mergers.

\section{The  star-formation history of the thin disk}

Unfortunately, the star-formation history (SFH) of the thin disk is
poorly known far from the solar neighborhood. This lack of data needs
to be rectified; hopefully the next generation of imaging surveys,
such as Pan-STARRS, will provide data for both the inner disk and the
outer disk.  Theoretical expectation in a wide range of models is that
star formation should proceed on faster timescales in the denser
regions, due to the shorter dynamical times there, so that inner
(denser) regions of disks are expected to have an older mean age, even
if the time of the onset of star formation is fixed.  Chemical
evolution models (see Pipino \& Matteucci's contribution to this
volume) also favor slower star-formation in the outer parts, to
match metallicity gradients.  In hierarchical-clustering models, as
noted above, the onset of star formation in the thin disk is later for
the outer parts, due to the later accretion of the higher-angular
momentum gas to form the outer disk.

\subsection{The onset of star formation in the local disk}

The star-formation history of the local disk has been derived using a
variety of techniques, the details of which are discussed (with
limitations and advantages) elsewhere in this volume.  Ages of the
oldest stars have been estimated from isochrone-fitting, with the
result that the oldest stars are less than 2~Gyr younger than the
metal-poor globular clusters, i.e.~ages of $\sim 11$~Gyr
(e.g.~analyses of the Hipparcos dataset by Binney et al.~2000; analyses of local stars with Str\"omgren photometry by
Nordstr\"om et al.~2004; Nordstr\"om, this volume; Holmberg,
Nordstr\"om \& Andersen 2008).  White-dwarf cooling ages are
model-dependent, as discussed at length in the contributions by
Salaris and by Kalirai in this volume, and oldest ages of $\sim
12$~Gyr are compatible with the luminosity function data (e.g.~Fig 8
of Salaris, this volume, but note that the models of Hansen et al.~2002, utilised
in Kalirai's paper in this volume, favour a younger oldest age).  These old ages are
consistent with an early onset of star formation in the local disk
(assuming that the stars found locally were born locally), the
lookback time of the onset corresponding to redshift $ z \simgt 2$.

The exponential scale-length of low-mass stars (of spectral type like the
Sun and later) in the disk is $\sim 2 - 3$~kpc (e.g.~Juri\'c et al.~2008
who used M-dwarfs as tracers). Thus if the old
stars in the solar neighborhood were formed close to their present location in the disk, star
formation was initiated at $\sim 3-4$ scalelengths at $z > 2$.  This
would then imply that the formation of extended disks was {\it not\/}
delayed until after a redshift of unity -- the typical epoch of the
last major merger for a $10^{12}\, M_\odot$ halo, in $\Lambda$CDM --
as has been proposed in CDM-models with feedback (e.g.~Weil et
al.~1998; Thacker \& Couchman 2001; Governato et al.~2007).

Alternatively, the old stars in the local thin disk could have formed
elsewhere and more recently arrived in the solar neighborhood -- two
such scenarios have been proposed, with these old stars forming in
either (a) satellite galaxies that are assimilated later, on circular
orbits (Abadi et al.~2003), or (b) the inner disk and migrating
outwards due to the influence of transient spiral arms\footnote{I
thank Roelf de~Jong for his question after my talk.}  (Ro\v{s}kar  et
al.~2008a,b).  In the first scenario, the satellites that could provide
stars on near-circular orbits at the solar neighborhood would most
probably have to be massive, so that dynamical friction (operating on
a timescale proportional to the inverse of the satellite mass) can be
effective in damping the satellite's orbit prior to its member stars
being accreted. These satellites are the most capable of
self-enrichment, and may be expected to contribute not just old stars,
but also younger stars.  If the pattern of elemental abundances
produced were anything like those found in the surviving satellites
(bearing in mind that these \lq old disk' satellites are proposed to
be accreted relatively recently, after a redshift of unity) then these
should give a distinct signature, in particular low values of
[$\alpha$/Fe] at low [Fe/H]. At least two groups -- Ruchti, Fulbright,
Wyse et al. (2009, in prep.), using the RAVE survey (e.g.~Steinmetz et
al.~2006) to select their sample, and Reddy \& Lambert (2008) -- are
obtaining and analyzing elemental abundance data for metal-poor
(thick) disk stars, and have found nothing distinctive.

In the second scenario, one again expects a signature in the elemental
abundance pattern of local stars, since efficient radial mixing of
stars from distant regions with different star-formation histories
gives increased scatter (e.g.~Fran\c{c}ois \& Matteucci 1993;
Schoenrich \& Binney 2009).  Migration has been proposed to explain in
particular metal-poor disk stars locally in the disk (Haywood 2008)
with a large fraction of these stars to have come inwards to the solar
neighborhood, from the outer disk.  These metal-poor disk stars are again of all ages
(except younger that $\sim 2$~Gyr, plausibly the travel time of the
migration), including the oldest ages, so this migration would imply an early
onset for the outer disk beyond the solar circle, even more difficult
to explain in $\Lambda$CDM than early star formation at the solar
circle.  Haywood (2008) proposes that migrated stars do indeed
complicate the interpretation of trends in the elemental abundance
patterns, but as we discuss below (section 3.2, cf.~Reddy, Lambert \& Allende
Prieto 2006), the scatter is small and the uncertainties in the kinematic basis for the assignment of stars to the
different components of the Galaxy provides an alternative explanation
for outliers.

\subsection {Variation over time}

Again, various techniques have been used to estimate the temporal
variation of the star-formation rate in the local disk. Several find
evidence for \lq bursts' in star-formation activity, of amplitude $2 -
3$, superposed on an underlying slow variation.  Examples include the
isochrone-based analysis of the Color Magnitude Diagram (CMD) of the
Hipparcos dataset by Hernandez, Gilmore \& Valls-Gabaud (2000),
providing a temporal resolution of 50~Myr, albeit with a sample
selection that limited the analysis to only the last $\sim 3$~Gyr (and
adopting a fixed metallicity, which is a reasonable simplification for
this narrow age range).  These authors found a quasi-periodic
variation with period of $\sim 0.5$~Gyr, which they suggested could be
due to the passage of spiral arms.  

Cignoni et al.~(2006) developed
their own approach to the analysis of the Hipparcos CMD and derived
the star-formation history  back to $\sim 12$~Gyr (adopting an
age-metallicity relation).  They found good agreement with the earlier
results of Hernandez et al.~for the younger stars, after
rebinning them into 1~Gyr bins to match  time
resolutions.  As is well-known, older disk stars have higher-amplitude
random motions, with resultant wider epicyclic excursions about their
orbital guiding-center.  The derived star-formation rates at older
ages then trace a larger portion of the disk, and stars formed locally
a long time ago may have been lost from the sample.  Cignoni et
al.~investigated the possible kinematic dependence of their derived
SFH by excluding stars more than $n \sigma$ ($n = 1,\, 2$) away from
the canonical local thin-disk 3D-space motion distribution, and found
that the resulting distributions, focussing on ages less than $\sim
6$~Gyr, were indistinguishable, and thus unaffected by dynamical
diffusion.

The major result is a broad peak at ages $\sim 2 - 6$~Gyr, with the
star formation rate at $\sim 3$~Gyr being a factor of 2-3 higher than
either of the present-day rate or the rate at ages greater than 6~Gyr.  This
increase is perhaps attributable to triggering of star-formation
activity by an accretion event.  This overall SFH is consistent with
the analysis of low-mass M-stars (using H$\alpha$ activity as an age 
indicator) by Fuchs, Jahrei{\ss}  \& Flynn (2009; see their Fig.~5 for
comparisons with others); the chromospheric-activity based analysis of
G-stars by Rocha-Pinto et al.~(2000) shows more high-frequency
variations and less of a consistent increase in star-formation
activity back to $\sim 5$~Gyr.

Fuchs et al.~(2009) argue that the Milky Way follows the {\lq}Schmidt-Kennicutt' star-formation law found for external disk
galaxies, which would imply that an increase in gas supply (inflow?
accretion?) accompanied the increased level of star-formation rate $\sim 5$~Gyr
ago. In any case it is reassuring to find that the Milky Way is
typical.
 
The possibility of significant re-distribution of stars via \lq radial
migration', in addition to the radial epicyclic excursions considered
already in the above papers, needs to be considered, as it raises the
issue of to which region(s) of the disk does the derived SFH apply?
Ro\v{s}kar et al.~(2008a) appeal to radial migration in particular to
build-up the outer disk, defined by a break in the gas surface density
(and correspondingly, following the Schmidt-Kennicutt law, in the
stellar surface density).  The outer HI disk of the Milky Way does
indeed show such a break, at Galactocentric distance of $12 - 
13$~kpc, well beyond the solar circle (see Fig.~5 of Levine, Blitz \&
Heiles 2008), and similar in location to the \lq edge' of the stellar
disk at $12 - 14$~kpc (Reyl\'e et al.~2008).  In the model of
Ro\v{s}kar et al.~(2008a), stars currently beyond the break are those
that are most likely to have migrated several kpc outwards ($\Delta R \sim
4$~kpc, see their Fig.~2), a significantly larger distance than their
typical epicyclic excursions of $\sim 2$~kpc.  The root-mean-square
change in radius across the scale, and lifetime, of all the disk stars
in this model is $\sim 2.4$~kpc, more comparable to the $\sim \pm
1.5$~kpc epicyclic excursions (estimated from observed kinematics) of the $\sim 3-5$~Gyr-old stars that
dominate locally.  Looking more closely
at the model `solar circle', Ro\v{s}kar et al.~(2008b) find that as
much as half of the stars could have been born outside $7-9$~kpc, with a
bias to metal-rich stars from interior regions.  As discussed above,
the mixing of stars from regions of different star-formation histories
should lead to scatter in the elemental abundances, and the evidence
is that only a small fraction of stars do not follow well-defined
trends, within observational uncertainty (Bensby et al.~2005; Feltzing
\& Bensby, this volume; Haywood 2008; Reddy, Lambert \& Allende Prieto
2006; Reddy \& Lambert 2008).  The implication is that radial
excursions are limited to mixing predominantly regions of very similar
star-formation histories and chemical enrichments -- either because
there is little radial variation in star-formation history over much
of the disk, or there is little radial migration.  It will be very
interesting to analyse the significantly larger samples of fainter stars with
elemental abundances that will be feasible with planned instruments
such as HERMES on the Anglo-Australian telescope, and WFMOS (a Gemini
instrument) on the Subaru telescope.

\section{The  star-formation history of the thick disk}

 \subsection{Formation and early evolution}

The Galactic thick disk was defined 25~years ago through star counts
at the South Galactic Pole in which two vertical exponential
components were manifest (Gilmore \& Reid 1983), with the general
concensus now that is separate and distinct from the Galactic thin
disk.  Analysis (Juri\'c et al.~2008) of the deep, uniform wide-field
imaging data from the Sloan Digital Sky Survey (SDSS) confirmed the
necessity for two disk components, and the best-fit `global'
thick-disk parameters are an exponential scale-length of 3.6~kpc and
exponential scale-height of 900pc, with 12\% of the local stellar mass-density in the thick disk. These combine to give a total mass equal to
10--20\% of the stellar mass of the thin disk, or $\sim
10^{10}\, M_\odot$.  Stars in the thick disk have distinct kinematics,
elemental abundances (and ratios) and age distributions, when compared
to the thin disk or stellar halo.  Similar structures have been
identified in the resolved stellar populations of external disk
galaxies (e.g.~Mould 2005; Yoachim \& Dalcanton 2006).
 
There are several mechanisms by which a thick stellar disk could
 result (see, e.g., Gilmore, Wyse \& Kuijken 1998; Majewski 1993).
 Thick disks seems inevitable in $\Lambda$CDM due to the heating
 inherent during the expected late merging and assimilation of
 satellites into pre-existing thin stellar disks.  Simulations with a
 cosmological distribution of subhalos, in terms of both their mass
 function and their orbital characteristics, confirm this, and also
 show that the most massive satellite dominates the heating
 (e.g.~Hayashi \& Chiba 2006; Kazantzidis et al.~2008; see Hopkins et
 al.~2008 for a dissenting view, based however on an analysis using the same
 satellite orbital distribution as Hayashi \& Chiba).  In agreement
 with earlier simulations that focussed on the accretion of one
 satellite in isolation, a satellite that is dense enough to survive to
 influence the disk, and of total dissipationless mass ratio 10-20\%
 of that of the stellar disk, will produce a thick stellar disk. 

 The heating is achieved {\it via\/} a mix of local deposition of
 energy plus excitation of resonances (Sellwood, Nelson \& Tremaine
 1998), and a thin-disk component can persist (Kazantzidis et
 al.~2008).  Of course, subsequent accretion of gas, and perhaps
 stars, can (re)-form a thin disk.  Dissipation naturally leads to a
 thin disk after accretion of gas, while accretion of stars into a
 thin-disk component requires circular orbits for the parent system of
 those stars.
 
The stars in the thick disk in this scenario (creation by heating from
a pre-existing thin disk) would have an age distribution that
reflected that of the thin disk that was heated into the thick
disk. With the continuous, fairly smooth, derived star-formation
history of the (local) thin disk, starting at the earliest epochs,
(lookback times $\sim 10-12$~Gyr), recent accretion, merging and
associated heating would then produce a thick disk that at the present
time would have stars of age equal to the look-back time of the
accretion, unless the accreted satellite were easily destroyed
e.g.~was of low (relative) density.  Turning this around, if the thick
disk originated through merger-induced heating of the thin stellar
disk, the last significant (defined as $> 20$\% mass ratio to the disk, robust, dense
satellite of stars and dark matter) merger can be dated by the young
limit of the age distribution of stars in the thick disk: an age
distribution of stars in the thick disk that goes down to, say, 5~Gyr
would allow a merger and heating at a redshift of $z_{last} \sim 0.5$,
when the lookback time equals 5~Gyr, but if all thick disk stars are
old, then the last significant merger was long ago.  The age of the
oldest thick disk stars, in this scenario, also further constrains
the epoch at which an extended, thin stellar disk was in place, available
to be heated.

\subsection{Ages of the oldest stars}

Analyses of the turn-off age of the thick disk stellar population, within a few kpc of the
solar neighborhood, agree that the bulk of the thick disk stars are
{\it old}.  The (well-defined) turn-off color, for the spectroscopically derived
typical metallicity of a star in the thick disk of $\sim -0.6$~dex, is equal to that of
Galactic globular clusters of similar metallicity, e.g.~47~Tuc,
corresponding to an age of $10 - 12$~Gyr (e.g.~Gilmore \& Wyse 1985; Carney,
Latham \& Laird 1989; Gilmore, Wyse \& Jones 1995; Ivezic et
al.~2008, their appendix).  This is equal to the estimated age of the
oldest {\it thin\/} disk stars locally, as discussed above.

Ages for {\it individual,} slightly evolved thick-disk stars can be
estimated from Str\"omgren photometry.  These analyses show that the
mean age is certainly older than that of the local thin disk, but
there are disagreements in the fraction of `thick disk' stars that are
younger than the globular cluster ages, plausibly largely due to the
difficulties in the assignment of an individual nearby star to a specific
component, either the thin or thick disk.  This probabilistic
assignment is based on kinematics, and the standard assumption is that
the space motions of each of the local thin disk and thick disk are
adequately modelled by three one-dimensional Gaussians (e.g.~see
Feltzing \& Bensby's contribution to this volume; Bensby et al.~2007a, 2004a,b;
Reddy, Lambert \& Allende Prieto 2006).  This is clearly an
over-simplification for the thin disk, given that moving groups have
been robustly identified in the local disk (e.g.~Dehnen 1998; Dehnen
\& Binney 1998; Famaey et al.~2005; Bensby et al.~2007b), with
properties consistent with being dynamically induced by a combination
of the Galactic bar and transient spiral arms (e.g.~Dehnen 2000; de
Simone, Wu \& Tremaine 2004).  As noted above, radial migration within the thin disk may occur and may also be a
complicating issue for the kinematics-based population assignments, and this has yet to be analysed in detail. 
The likely production of high-velocity
outliers in the thin disk kinematics by such mechanisms as three-body
interactions can also complicate the population assignments. 

Indeed,
Reddy et al.~(2006; their Fig.~24) suggest that a significant fraction
of `younger' thick disk stars are in fact contaminants from the thin
disk.  Evidence for an age-metallicity trend within the thick disk
does remain in their sample, although once these thin disk
contaminants are removed, the thick disk shows no evidence for the
incorporation of iron from Type~Ia supernovae, in the elemental
abundance pattern (their Fig.~20).  An age spread of several Gyr
could be consistent with this constant `Type~II plateau' in the
[$\alpha$/Fe] ratios, plus a typical timescale for Type~Ia chemical
enrichment of $\sim 1$~Gyr, if the thick disk consisted of stars from
several independent star-formation regions, each of which had a short
($\simlt 1$~Gyr) duration, but different onset times.  However, with
age uncertainties of $\pm 2$~Gyr, the data are also consistent with a
narrow age range, with mean $\sim 12$~Gyr (on their age-scale), and a simpler
interpretation of the elemental abundance pattern as reflecting a global short duration of star formation, $\sim 1$~Gyr.
 
\subsection{Implications for merger history}

The old age of thick disk stars, $\simgt 10$~Gyr, limits the last
significant minor merger to have occured at a redshift $ \simgt
2$. While the last {\it major\/} merger in $\Lambda$CDM typically is
at this epoch, it is minor mergers that are constrained by heating of
the thin disk into a thick disk, and these are expected to continue to
lower redshifts (e.g.~Stewart et al.~2008).  The inferred quiescent
merger history of the Milky Way is atypical in $\Lambda$CDM.
As we will now discuss, 
consistent limits on the minor-merger history are obtained from the  
derived star-formation history of the  central bulge.

\section{The  star-formation history of the central bulge}

During mergers, the expectation is that existing disk stars, and gas to
fuel star formation, will be added to the bulge (e.g.~Kauffmann 1996),
perhaps through an intermediate stage of build-up of a massive inner
disk that subsequently becomes unstable (e.g.~Bower et al.~2006).  The
dense, inner regions of satellites can also survive to be added to the
bulge, if dynamical friction is efficient enough (and the satellite
massive and dense enough). Gravitational torques due to the (mildly)
triaxial inner bulge/bar will also drive modest gas inflows at the
present day (e.g.~Englmaier \& Gerhard 1998), in the plane of the
disk. 
 
As discussed in detail in Fulbright's contribution to this volume,
while there are younger stars in the central regions, these are
confined to the disk plane, and the central bulge is dominated by old
($10 - 12$~Gyr), metal-rich stars, with enhanced alpha-element ratios.
These properties point to bulge formation in an intense short-lived
burst of star formation, {\it in situ\/} (a deep potential-well being
required to reach the observed high metallicities), a long time ago
(e.g.~Elmegreen 1999; Ferreras, Wyse \& Silk 2003; and the contribution
by Pipino \& Matteucci in this volume). 
The inferred star-formation rate is reasonable, of order 10~$M_\odot$/yr for a total mass of $\sim 10^{10}\, M_\odot$ and an age range of $\sim 1$~Gyr. 

Thus unless the inner disk is composed of uniformly old stars, there
can have been no recent disk instability to form the bar/bulge. There
is little room for significant build-up of the central bulge by recent
mergers, with the old age again limiting significant merger activity
to redshifts $\simgt 2$.  This matches the constraints from
the thick disk, and leads to the suggestion (Wyse 2001) that perhaps both bulge
and thick disk were formed by the same last significant minor merger.

An alternative suggestion, consistent with the low value of the angular momentum content of
the bulge, and the similarity of the specific angular momentum distributions of
the bulge and stellar halo (see Fig.~1), is that the bulge formed from gas ejected
from the early star-formation regions in the halo (Carney, Latham \& Laird 1990; Wyse \& Gilmore 1992).  Gas must have been lost from the stellar halo since all indications are that the stellar Initial Mass Function (IMF) was normal, at both high and low masses (e.g.~Wyse 1998), but the mean metallicity is far below the yield for that IMF (see Hartwick 1976 for the basic model).

 \begin{figure}[h!]
\begin{center}
 \includegraphics[angle=0,width=3.5in]{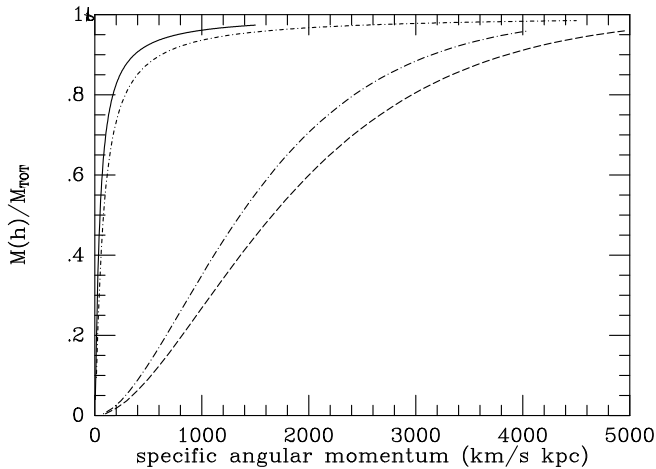}
\caption{Adapted from Wyse \& Gilmore 1992, their Figure~1.  Specific angular
momentum distributions of the bulge (solid curve), the stellar halo
(short-dashed/dotted curve), the thick disk (long-dashed/dotted curve)
and the thin disk (long-dashed curve).  The bulge and stellar halo
have similar distributions, with most of the mass at low angular momentum (curves passing through upper left of the figure). The distributions of thick and thin disks are also similar to each other, with significant mass fraction at high angular momentum.}
\end{center}
\end{figure}

\section{The  star-formation history of the halo}

The bulk of the stellar halo by mass, interior to Galactocentric
distances of $\sim 20$~kpc, is rather uniform in its properties: the
stars -- and globular clusters -- are old and metal-poor, and show enhanced elemental abundances
([$\alpha$/Fe] in particular) that are indicative of short duration(s)
of star formation, in low-mass star-forming regions, with `normal' 
Type~II-progenitor stellar IMF.  These properties
are unlike those of most stars in satellite galaxies now.  Particular
attention recently has been paid to the differences in elemental
abundance patterns (cf. Venn et al.~2004; Geisler et al.~2007). The
differences can be understood in terms of the different star-formation
histories of the field halo and the surviving satellite galaxies --
the former having a narrow stellar age-range and rapid chemical enrichment,
and the latter having a wide stellar age-range and slow, inefficient  enrichment, allowing
incorporation of iron from Type~Ia supernovae (and hence low [$\alpha$/Fe]) even at low levels of
overall enrichment (e.g.~Gilmore \& Wyse 1991; Unavane, Wyse \&
Gilmore 1996; Lanfranchi \& Matteucci 2003). 
 
Comparison of the age distributions alone leads to the conclusion
(Unavane et al.~1996) that accretion into the field halo from stellar
satellites with a typical extended SFH has not been important for the
last $\sim 8$~Gyr, this time coming from the estimated lower limit to
the ages of stars in the field halo.  Their analysis of the distribution of the field
halo stars in the color-metallicity plane allows perhaps $\sim 10$\%
by mass to have been accreted later, predominantly in the metal-rich
tail of the halo.  Of course satellites like the Ursa Minor dSph,
composed of only old, metal-poor stars, could be accreted and
assimilated into the field halo at any time and would not be
distinguishable on the basis of age or metallicity.  However, such
satellites are rare at the present time.  Early disruption of a few massive  
satellites could form the stellar halo, of total stellar mass $\sim 10^9\, M_\odot$ (e.g.~Robertson et al.~2005), but we need to understand what causes the necessary cessation in star formation, when, for example, the LMC has not been so affected. 

A significant fraction of stars in the outer halo could have been
accreted from the Sgr dSph (Ibata, Gilmore \& Irwin 1994), which is on
an orbit with periGalacticon of $\sim 25$~kpc and apoGalacticon $\sim
50$~kpc (e.g.~Ibata et al.~1997). Tidal arms from this system are seen across the Galaxy (e.g.~Majewski et al.~2003; Fellhauer et al.~2006).

\section {Concluding remarks}

The old mean stellar ages and short durations of star formation of the
thick disk and bulge argue for little late accretion and merging into
the Milky Way.  Accretion after a redshift of $\sim 2$ should have
been predominantly smooth, of gas-dominated, low-density systems.  The
relatively high mean metallicity, plus inferred rapid star formation
(from the ages and elemental abundance patterns), of both components 
argues for star formation within deep potential wells.  This favours
{\it in situ\/} star formation of each of the bulge and thick disk,
and rapid mass assembly of the overall Milky Way.
 
The presence of very old stars, ages $\simgt 10$~Gyr, in the local
thin and thick disks argues for the existence of an extended disk at
redshift $\sim 2$.  

The old mean stellar age and inferred short duration of star formation
of the bulk of the stellar halo implies that the field halo cannot
have formed from systems like the existing satellites, which typically
have had much more extended star-formation histories.  The low mean
metallicity, plus curtailed star formation, argues for star formation
within shallow potential wells, so that early mass loss is
facilitated.  Instead of such low-mass systems, the $\Lambda$CDM-based
models favour a few massive satellites, accreted early, as the source
of the field halo (Robertson et al.~2005; Sales et al.~2007); rapid
gas loss is assumed to occur by ram pressure stripping, but this assumption needs to be modelled.

 The quiescent merging history, and rapid mass assembly, of the Milky Way, is unusual in $\Lambda$CDM.  However,  surveys of     
galaxies at redshift $z \sim  2$, find rapid star    
formation and chemical enrichment, and   
extended disks, similar to the inferences for the Milky Way at that look-back time  (e.g.~Maiolino et al.~2008;  Daddi et al.~2008; Genzel et al.~2008).

However, our understanding of the global star-formation history  of the Milky Way remains
incomplete.  For this we need to determine the detailed age distributions,
spatial distributions, space motions, and elemental abundances for
large samples of Galactic stars, both locally and more globally.
Several large imaging surveys are planned in the near future, and
these need to be matched by large spectroscopic surveys, at both high
and low spectral resolution. These will be challenging, but worth the investment.

\newpage
\begin{discussion}

\discuss {de Jong}  {Recently it has, for instance, been argued by Ro\v{s}kar  
et al.~(2008) that radial migration of stars is much larger than  
previously estimated, meaning that stars in the solar neighbourhood do  
not reflect the local star formation history, but instead a  
combination of central Galaxy star formation and migration history.   
If this is the case, how can we disentangle these effects?}

\discuss {Wyse}  {The thin disk at the solar neighborhood shows little  
scatter in elemental abundance ratios, and this is hard to reconcile  
with stellar migration over many kpc, mixing regions of different  
star-formation history.  If migration is limited to $\simlt 1$~kpc then I  
do not believe it will affect the derived local SFH significantly 
since normal stellar orbits  sample that range.}

\discuss {Melbourne}  {What percentage of stars in the thin disk are  
old?  Are they coeval with the thick disk?  Would they be thick disk  
stars?}

\discuss {Wyse} {The analyses of the derived local star-formation
histories do not have a very good handle on the oldest stars (since
they are faint) but estimates (e.g.~Cignoni et al.~2006) suggest
$\sim$10\% in the 10-12 Gyr range.  These have thin-disk kinematics so
they are probably not thick-disk members (modulo uncertainties in population
assignment). It is difficult to distinguish ages 10-12 Gyr, so
although the oldest thin disk stars may be $\sim 1$~Gyr younger than
the thick disk, it is best to say both are `old'.}

\discuss {King}   {You've said little or nothing about the bar.  Is it  
merely the manifestation of a dynamical instability, or should we be  
able to learn something from it about the formation history of the  
Milky Way?}

\discuss{Wyse}{We still don't know much about the evolutionary state  
of the bar.  It is clear that the Milky Way bulge shows some  
characteristics of `pseudo-bulges', for example an exponential  
surface-density profile, but the uniform old age argues against a  
recent disk-bar instability to form the bulge, given the evidence for  
recent star formation in the inner disk.  But we do need to get more  
data on the stellar populations in the inner disk.}

\end{discussion}

\end{document}